\documentclass[sigchi]{acmart}

\copyrightyear{2023}
\acmYear{2023}
\setcopyright{rightsretained}
\acmConference[CHI '23]{Proceedings of the 2023 CHI Conference on Human Factors in Computing Systems}{April 23--28, 2023}{Hamburg, Germany}
\acmBooktitle{Proceedings of the 2023 CHI Conference on Human Factors in Computing Systems (CHI '23), April 23--28, 2023, Hamburg, Germany}\acmDOI{10.1145/3544548.3581328}
\acmISBN{978-1-4503-9421-5/23/04}

\usepackage{enumitem}
\usepackage{csquotes}
\usepackage{xcolor}

\begin{document}


\newcommand{\iquote}[1]{\textit{``#1''} }

\definecolor{royalazure}{rgb}{0.0, 0.22, 0.66}

\newcommand{\todo}[1]{{\textcolor{red}{[\textit{#1}]}}}
\newcommand{\pgk}[1]{{\textcolor{violet}{[\textit{#1} --PGK]}}}
\newcommand{\sunny}[1]{{\textcolor{cyan}{[\textit{#1} --SC]}}}
\newcommand{\bethg}[1]{{\textcolor{blue}{[\textit{#1} --BG]}}}
\newcommand{\ian}[1]{{\textcolor{orange}{[\textit{#1} --IB]}}}
\newcommand{\dan}[1]{{\textcolor{magenta}{[\textit{#1} --DR]}}}
\newcommand{\mia}[1]{{\textcolor{teal}{[\textit{#1} --MH]}}}

\newcommand{\eg}{e.g., }
\newcommand{\etc}{etc.}
\newcommand{\ie}{i.e., }
\newcommand{\etal}{et al.\@\xspace}

\title[Practicing Information Sensibility]
{Practicing Information Sensibility: \\
How Gen Z Engages with Online Information}

\author{Amelia Hassoun}
\affiliation{\institution{University of Cambridge}\country{United Kingdom}}
\email{ah2229@cam.ac.uk}

\author{Ian Beacock}
\affiliation{\institution{Gemic}\country{Canada}}
\email{ian.beacock@gemic.com}

\author{Sunny Consolvo}
\affiliation{\institution{Google}\country{USA}}
\email{sconsolvo@google.com}

\author{Beth Goldberg}
\affiliation{\institution{Jigsaw}\country{USA}}
\email{bethgoldberg@google.com}

\author{Patrick Gage Kelley}
\affiliation{\institution{Google}\country{USA}}
\email{patrickgage@acm.org}

\author{Daniel M. Russell}
\affiliation{\institution{Google}\country{USA}}
\email{dmrussell@gmail.com}


\renewcommand{\shortauthors}{Hassoun et al.}

\begin{abstract}
Assessing the trustworthiness of information online is complicated. Literacy-based paradigms are both widely used to help and widely critiqued. We conducted a study with 35 Gen Zers from across the U.S. to understand how they assess information online. We found that they tended to encounter---rather than search for---information, and that those encounters were shaped more by social motivations than by truth-seeking queries. For them, information processing is fundamentally a social practice. Gen Zers interpreted online information together, as aspirational members of social groups. Our participants sought \textit{information sensibility}: a socially-informed awareness of the value of information encountered online. We outline key challenges they faced and practices they used to make sense of information. Our findings suggest that like their information sensibility practices, solutions and strategies to address misinformation should be embedded in social contexts online. 

\end{abstract}


\keywords{information literacy, information sensibility, teens, youth, Gen Z, search, social media, misinformation, disinformation, rumors}

\maketitle

\section{Introduction}

Assessing the trustworthiness of information online is an increasingly complicated task in the current era of information abundance. Schools, governments, and online platforms use literacy-based paradigms\footnote{ We use the term ``literacy-based paradigms'' to describe a range of normative, pedagogical paradigms that define and measure core individual competencies for assessing and processing different kinds of information (see Section \ref{sec-rlit}).}
in an attempt to tackle the growing problem of online misinformation (\eg~\cite{unesco2013global, breakstone2022civic, twitter2019, bulger2018promises, garcia2013transforming}). From their earliest origins in the mid-20th-century fight against propaganda \cite{hobbs2014teaching}, such approaches have been linked to civic education for democracy and the public interest \cite{mihailidis2013media}. Despite significant investments in individual information literacy skills, however, people remain vulnerable to misinformation online \cite{livingstone2019global}, and such literacy-based paradigms have been widely critiqued in academic literature \cite{boyd2018you}. 

Against the backdrop of this information crisis, we sought to understand how members of Gen Z\footnote{We follow the Pew Research Center definition of Gen Z as people born between 1997 and 2012 (i.e., who were between the ages of 10 and 25 during our data collection) \cite{dimock2019defining}.}
(``Gen Zers'') engage with information online. To do so, we performed a qualitative study with 35 Gen Z internet users to investigate: \textbf{RQ1 - Seeking information} --- Why, where, and how do Gen Zers seek information online? \textbf{RQ2: Assessing information} --- Under what circumstances do Gen Zers attempt to assess if information they engage with online is accurate, credible, and high-quality? How do they perform such assessments? \textbf{RQ3: Susceptibility to misinformation} --- What factors affect the susceptibility of Gen Zers to misinformation?


We found that our participants understand \textit{information} as referring to more than facts or news. To them, it means a much wider array of stories, statements, facts, attitudes, beliefs, and commentary from institutional and personally relatable sources. This kind of information is found in different media formats including text, images, videos, and audio. We follow Bates' \cite{bates2005info} second definition of information as patterns of organization which have been given meaning by a living being, focusing our analytical attention on the emergent meaning attributed by our participants to these multi-media messages. We adopt a shared information behavior (IB) and human-computer interaction (HCI) approach to studying information-as-process, which "includes all aspects of needing, seeking, searching for, encountering, and using information," expressed in terms of "sense-making" and "meaning-making" \cite{gorichanaz2022info,bates2017encyclopedia}. We intentionally did not limit the type of information explored, as it allowed us to understand how participants themselves parsed and categorized information -- practices which, as we detail, were often not well-described by existing frameworks. 

We initially set out to investigate literacy-based frameworks among Gen Zers with questions like: What are the core competencies for evaluating online information? Can Gen Zers recognize logical fallacies and biases? How can we help them be better researchers and fact-checkers to protect them against misinformation? Yet we found that these questions led us to focus on individual strengths and weaknesses, occluding the factors that most affected our participants.

Our Gen Z participants repeatedly expressed more socially-oriented concerns. When engaging with online information, they asked questions like: What would my parents think? Do I feel a personal connection with this source or influencer? Is this information based on first-hand experience? What do people ``like me'' think about it? Why are so many people sharing it? How could it help me establish who I am? For them, information-seeking was motivated by more than a desire for fact or truth; social motivations shaped how and why Gen Zers sought and interpreted information. Social motivations also helped them evaluate the significance and relevance of information---a process in which credibility and accuracy were only one part of the equation. Rather than information literacy, our participants sought what we term \textit{information sensibility}: a socially-informed awareness of the value of information encountered online, relying on folk heuristics \cite{wash2010folk, d2005some} of credibility. While other work has cited the role of social influence on misinformation evaluation, we argue it is not just that social ties affect users when they evaluate online information---social belonging itself is often the primary motivation for (mis)information engagement.


In order for the HCI community to develop solutions that effectively equip Gen Z---and potentially others---with strategies to identify misinformation, we argue that a contextual understanding of their information sensibility practices is needed \cite{saltz2021encounters}. Our research yielded 3 key implications: 
\begin{enumerate}
	\item	Gen Zers' informational and social needs are inseparably entangled.
    \item	Gen Zers’ information journeys often do not begin with a truth-seeking search query.
    \item	Gen Zers use information to orient themselves socially and define their emerging identities.
\end{enumerate}
We begin by situating our work at the intersection of literatures about online information, information literacy, and Gen Zers' online behavior. We describe our methodology, then transition to results, outlining challenges participants faced. We present four practices participants used to make sense of information and the trust heuristics underlying them. Finally, we discuss which of these challenges, practices, and heuristics may increase susceptibility to misinformation---and how an information sensibility framework could help interventions better reach Gen Zers. We also hypothesize which of our findings are period, age, and cohort effects, suggesting where findings may have wider implications to guide future research. We conclude that since Gen Zers use information to orient themselves socially, solutions and strategies to reduce their vulnerability to misinformation should also be socially embedded.


\section{Related Work}\label{sec-rw}

This paper reframes discussions about Gen Z susceptibility to misinformation in two ways. First, it shifts the focus from individuals to socially-positioned subjects to better account for the social motivations behind information-seeking and evaluation. Second, it adopts an inductive, ethnographic, and ecologically-valid methodological approach to questions that have often been studied deductively and/or experimentally. Our work most directly builds upon the HCI literature about online information-seeking and vulnerability to dis/misinformation, including literacy-based approaches to information evaluation. We also relate our research to an interdisciplinary literature on the digital practices of Gen Zers. 

\subsection{Seeking \& Evaluating Online Information}\label{sec-rlit}

This study adjusts the focus of prior HCI work on information-seeking by investigating the in situ social motivations that users bring to their engagement with online information, an underexplored, but critical, topic in a literature that emphasizes practices and behaviors. Our findings contribute to a body of work---shaped by early internet tools like search engines and desktop browsers---about how users seek and evaluate online information. This work often defines ``information'' as news~\cite{qayyum2010investigating, edgerly2017seeking, bentley2019understanding} or specialized facts (\eg health information~\cite{hargittai2012searching, jia2021online}) rather than interpretations or arguments, beliefs, and opinions. Areas of focus include web page affordances (\eg visualization of search results), user credibility judgments~\cite{wathen2002believe, hong2006influence,giudice2010crowd, westerwick2013effects, lurie2018investigating}, and offline social determinants (\eg age, race, education) shaping information-seeking practices~\cite{harrington2022s, ragnedda2022offline}. Although one 2019 study found only 16\% of desktop news-seeking journeys starting with social media~\cite{bentley2019understanding}, rising mobile and social media usage has remade the information landscape in ways we are only starting to understand. Our work validates much earlier findings from Evans and others that online searching is more social than often assumed~\cite{smith2008leveraging, evans2008towards, evans2010elaborated, evans2010your, eynon2012understanding, jeon2013value}, but it shifts attention from sociality of behavior to sociality of motivations. In other words: even as the social practices of internet search grow clearer---via an emerging body of work on social media searching behaviors~\cite{lampe2006face, kulshrestha2017quantifying, neely2021health, zhang2022shifting}, passive information exposure on social media~\cite{fletcher2018people}, and the role of online communities in enabling nonlinear information-seeking~\cite{patel2019feel, lee2021viral}---the varied and socially-informed motivations that users bring to information-seeking (beyond fact-finding or verification) remain understudied.

While we focus on the social influences that shape information-seeking, recent work into online misinformation and disinformation has stressed the social influences affecting the evaluation and circulation of misleading or low-quality information. From ``fake news'' to the COVID-era ``infodemic''~\cite{lewandowsky2012misinformation, flintham2018falling, zarocostas2020fight}, this work has convincingly demonstrated that such phenomena must be understood socially: from the distributed ``collaborative work'' that goes into producing mis/disinformation~\cite{starbird2019disinformation, lee2021viral} to users' use of social cues---like consensus and crowdsourcing heuristics~\cite{babaei2018purple, pennycook2019fighting,giudice2010crowd} or social endorsements~\cite{messing2014selective, geeng2020fake}---to make credibility judgments and decide whether to share information~\cite{lottridge2018let, herrero2020teens}. Scholars like Phillips \& Milner~\cite{phillips2021you} have applied this approach  further downstream, showing how effects of mis/disinformation are also intersubjective phenomena. Our work applies this social approach to online information generally, explicitly focusing on users' motivations, seeking behaviors, and evaluation practices in the context of their holistic information ecologies (rather than pre-defining and analyzing their evaluations of specific content like "news"). We argue that a contextual, socially-attuned sense of what users treat as information, why they engage with it, and what they use it for alters our understanding of misinformation susceptibility, as well as the interventions needed to support Gen Zers. 

Adjacent to this literature on how users engage with online information is a body of work about how they \textit{should} do so: identifying and evaluating the competencies needed to resist false or misleading information. Researchers have developed several frameworks grounded in the metaphor of \textit{language literacy}~\cite{lakoff1980metaphors,mason2021metaphor} which frame information consumption as a rules-based decoding process. The highest-order framework is often thought to be \textit{media literacy}---defined as a person’s ability to ``access, analyze, and produce information for specific outcomes''~\cite{aufderheide2018media}. Beneath it sit the technical skills of \textit{digital} and \textit{web literacy}~\cite{allmann2021rethinking,caulfield2017web,eshet2004digital,hobbs2010digital,martzoukou2020study} and the critical reading/research skills of \textit{news literacy}~\cite{klurfeld2014news,vraga2021news,chang2020news} and \textit{information literacy}~\cite{american1989presidential,de2020information,jones2021does}.  

In this work, however, we follow Mackey \& Jacobson~\cite{mackey2011reframing} in treating information literacy as the metaliteracy that captures three features common to all literacy-based approaches to engaging with online information. First, information literacies tacitly rely on an individual linear model of engagement and evaluation: a single consumer encountering a discrete piece of information and critically evaluating its quality and/or veracity. Second, these approaches assume that this can be done most effectively by mastering core competencies, from fact-checking to lateral reading (comparing one piece of information with other sources~\cite{breakstone2021lateral, wineburg2022lateral, caulfield2017web}). Finally, these paradigms assume that the most accurate information is also the highest-quality or highest-value information for any given individual, and that validating accuracy is the primary way in which people engage~\cite{marwick2018people}. Despite the dominance of information literacy, its weaknesses are well-rehearsed. Most salient for our work are critiques by scholars like Marwick and boyd (discussed further below), who note that information literacy relies on a flawed vision of human beings as rational individuals~\cite{boyd2017did, boyd2018you, phillips2019light, phillips2019toxins, phillips2021you, marwick2018people}.

Our work moves beyond these critiques by adopting a wider (and more ecologically-valid) definition of information than most literacy-based approaches, while focusing on the motivations and needs that help users find, engage with, and make sense of the information that is meaningful to them. In these ways, we aim to better account for the sociality of information-seeking and evaluation in a networked age of social media, from Reddit to TikTok. We do so by examining the cohort most associated with this information landscape: Generation Z.

\subsection{Digital Information Practices of Gen Zers}\label{sec-rprac}

By adopting an ethnographically-informed approach to Gen Zers' online practices rather than evaluating competencies, we suggest an alternative way of understanding---and reducing---Gen Zers' susceptibility to online misinformation. 

Although some sociological accounts \cite{katz2021gen} have offered positive interpretations of Gen Z and their digital capabilities, most---building on a tradition of skepticism about how the internet may cause social or psychological damage \cite{turkle2017alone}---argue that Gen Zers are risk-averse and safety-focused, unable to manage in-person social interactions, withdrawn from others, and unprepared for adulthood \cite{twenge2017igen, lukianoff2019coddling}. Optimists and pessimists agree, however, that Gen Zers do not recognize a sharp boundary between ``online'' and ``offline,'' highly value markers of identity and authenticity, face an unprecedented deluge of information, and feel simultaneously empowered and powerless \cite{katz2021gen, stahl2022genz}. 

Most work on Gen Zers' digital practices focuses on how they evaluate online information using the skills and competencies of relevant literacy frameworks (news, digital, media, information, etc.)\footnote{These findings emerged as a subset of a wider interdisciplinary discussion (\eg  HCI, media studies, sociology, political science, behavioral sciences) focused on reducing susceptibility to misinformation. Accordingly, many studies that are relevant for our purposes here do not explicitly adopt “Gen Z” as a category of inquiry, but instead focus on “adolescents,” “teenagers,” “youth” and/or “young people” whose birth years overlap with the cohort represented by Gen Z.}. This research is largely defined by the evaluation of predetermined behaviors and competencies (\eg \cite{chi2021analysis}), the performance of those competencies in controlled settings, and a focus on practices of online information-seeking (\eg web browsing, search-engines) established before the rise of social media \cite{breakstone2022civic}. 

Studies have found that despite high levels of confidence in their own digital abilities \cite{nygren2019swedish, porat2018measuring}, Gen Zers exhibit severe information literacy deficits and perform poorly at evaluating the credibility of online sources, finding accurate information, differentiating news stories from advertisements, and overcoming motivated reasoning to critically judge the accuracy of truth claims \cite{breakstone2022civic, breakstone2021students, mcgrew2018can, kahne2017educating, hargittai2012searching}. Gen Zers have often been found to deploy what one study \cite{list2016undergraduate} terms ``nonepistemic'' criteria when evaluating, selecting, and sharing online information (including relevancy, accessibility, social significance, and interactional elements like search-engine rankings), rather than strictly epistemic criteria, like credibility or reliability (\eg \cite{wolf2022saw, john2021, herrero2020teens, lurie2018investigating}). It is not clear whether these patterns are more pronounced for Gen Zers than earlier cohorts \cite{hargittai2010digital, pan2007google}. 

Gen Zers have most often been tested on their ``civic online reasoning'' abilities (\eg \cite{breakstone2022civic, breakstone2021students}), but similar deficits have been found when seeking health information \cite{hargittai2012searching} and content relating to race \cite{tynes2021google}. These deficits have also been identified in Gen Zers outside the U.S. \cite{gui2011digital, nygren2019swedish}. Pedagogical interventions have been shown to improve the ability of Gen Zers to judge the credibility and evaluate the truth value of information online (\eg \cite{wineburg2022lateral, breakstone2021lateral, mcgrew2020learning} but cf. \cite{kohnen2020can}).

We join a recent shift to focus not on pedagogy but rather on practice: that is, how Gen Zers engage with information in their daily lives. The COVID-19 pandemic revealed, for instance, that Gen Zers exhibit strong levels of digital and information literacy when confronted with public health misinformation \cite{feng2022investigating, volkmer2021social}. A global study from the World Health Organization found Gen Zers using classic information literacy strategies plus novel tactics (like algorithmic manipulation or ``domestication'' \cite{simpson2022tame}), but they also reported skepticism, exhaustion, frustration, and withdrawal in response to the overwhelming volume of online information \cite{volkmer2021social}. 

Researchers have also found that Gen Zers are frequently familiar with literacy strategies but do not always use them, due to family pressures, social sanction, or judgments about relevancy or significance \cite{feng2022investigating, almeida2022does}. For Gen Zers, mockery or satire can mask a sophisticated understanding of information literacy \cite{vacca2022you, perovich2022self, literat2021research, lorenz2021birds}. As Almeida et al. note~\cite{almeida2022does}, the key question may not be about knowledge and its enactment, but rather: when does that knowledge matter for young people---and when does it not? We analyze this question throughout the rest of this paper.

In sum, understanding the social dynamics and motivations of engaging with information online is crucial to the study of misinformation susceptibility, especially for Gen Zers navigating social media in the age of TikTok. This study combines the inductive ethnographic approach taken by recent work on digital youth cultures with questions pioneered by HCI researchers interested in online search behaviors, credibility, and information literacy.  Our contribution thus lies in offering a rich, ecologically-valid account of the many social lives of online information~\cite{brown2000social}---what users believe to be meaningful information to begin with, how and with whom they seek it, and for what reasons young people choose to seek, metabolize, or give meaning to particular information---in order to improve our ability to design interventions that can effectively and sustainably reduce user vulnerability to misinformation. 
\section{Methodology}

We conducted a mixed-methods study in Spring 2022 with 35 Gen Z internet users aged 13--24 from the U.S. Here, we describe our participants, methods, ethical considerations, and study limitations. 

\subsection{Participants \& Recruiting}

We partnered with a market research firm to recruit Gen Z internet users from across the U.S. using typical recruitment methods (e.g., posting ads and invitations online, snowball recruiting, etc.). At the time of data collection (March and April 2022), participants were 13--17 years old ($n=17$) or 18--24 years old ($n=18$). Each U.S. region 
was represented: South (11), West (10), Midwest (7), and Northeast (7). Participants were from a mix of self-reported urban (9), suburban (22), and rural (4) areas, and spanned a range of education levels and self-identified political viewpoints (see Table~\ref{tab-partic}). 

We recruited a mix of mainstream ($n=19$) and alternative ($n=16$) media users\footnote{Eleven of the 13--17 year olds were classified as “mainstream” and 6 “alternative”; 8 of the 18--24 year olds were “mainstream” and 10 “alternative.” When classifying participants as "alternative," we acknowledge the challenges of self-reporting non-normative behavior and the role of parents in mediating responses from minors.} to explore whether platform use informed information literacy beliefs and practices\footnote{Alternative platforms were defined as congruent with the “alt-tech” ecosystem \cite{ebner2019alt-tech}. These included in-house “safe haven” platforms developed by far-right extremists; ultra-libertarian platforms; and alternative social forums popularized due to the perception that alternative hosts provide less stringent content moderation than mainstream platforms. To classify participants, we asked whether they used 4chan, 8kun, Gab, Parler, or Rumble three or more times/week. We then refined classifications based on their answers to: “Do you use any apps, platforms, sites, or channels that your friends don’t? What are they? Do you have any opinions or views that you would describe as alternative or non-mainstream? Are you part of any online communities that you don’t tell anyone about? Do you have any political opinions that you don’t feel comfortable sharing with others?” Four were reclassified, two in each direction.}. We recruited a set of participants representing half of our sample, and then asked them to invite someone they knew from within their age bucket to join the the study, completing our sample of 35 participants.\footnote{41 participants---including direct recruits and those who joined via friend or family invitations---started the study; 35 completed it. Data collected from the 6 who did not complete the study was deleted and not used in our analysis. The 6 non-completions were due to scheduling difficulties; no one formally requested to withdraw.} To maintain sample diversity, we tracked the details of participants who were invited by another participant to join the study against our quotas. 

\begin{table*}
    \centering
\begin{tabular}{lr @{\hskip 3em} lr}
\toprule
\bf Highest Level of Education Completed & \large $n$ & \bf Political Leaning (self-identified) & \large $n$ \\ \cmidrule(r{2.5em}){1-2}\cmidrule(l{4.0em}){2-4}
Some High School & 16 & Very Conservative & 2 \\
High School Diploma & 1 & Conservative & 4 \\
 Some College & 9 & Moderate & 14 \\
 2-Year Associate Degree & 4 & Liberal & 10 \\
 4-Year Bachelor’s Degree & 4 & Very Liberal & 5 \\ 
 Prefer to Not Say & 1 \\ \bottomrule
\end{tabular}
    \vspace{0.5em}
    \caption{Participants' education levels \& political leanings. Highest levels of education completed by and self-identified political leanings reported by the 35 participants.}
    \Description[Participants' education levels and political leanings]{Table 1 lists the highest levels of education completed by and political leanings reported by our 35 participants. Regarding education, 16 participants’ highest level of education completed was some high school, 1 participant had a high school diploma, 9 participants had some college, 4 participants had a 2-year Associate Degree, 4 participants had a 4-year Bachelor’s Degree, and 1 participant preferred not to say. Regarding self-identified political leaning, 2 participants reported leaning very conservative, 4 leaned conservative, 14 leaned moderate, 10 leaned liberal, and 5 leaned very liberal.}
    \label{tab-partic}
\end{table*}

\subsection{Research Methods}

Our methods were informed by the 
insight that people do not always do what they say---or say what they do \cite{rosaldo1993culture}. We sought to understand information literacy through our participants' perspectives, contextualized in their socioculturally-mediated values. 
Our study included three phases of data collection. Each participant provided informed consent before their first session.\footnote{For minors, informed consent was provided by parents or legal guardians in writing and verbally. The study was verbally explained to the minor. Researchers ensured that  parents/guardians and minors understood the study and reminded them that they could discontinue participation at any time without consequence. Informed consent was again obtained during the dyadic session, and participants were empowered to end the interview or redirect the conversation at any time.}

\paragraph{Phase 1 – Dyadic Interviews.} We began with interviews in which we asked pairs of participants who knew each other about their life histories, family values and influences, and the effects of social and economic context on their online behaviors, information-seeking practices, and trust heuristics.\footnote{We employed dyadic interviews at the suggestion of several experts---including those with experience conducting research with minors---consulted during study planning.} During these sessions, participants shared their screens as they demonstrated information-seeking. These methods surfaced participants' attitudes around information literacy as well as their actual behaviors. They also helped us uncover meaningful differences between reported and observed behaviors, which we unpacked with the participants. 

Dyad partners---often siblings, romantic partners, or friends---were from the same age bucket; some were a mix of mainstream and alternative media users. In three cases, they were from different U.S. regions. Each session was conducted remotely and lasted 90 minutes. We collected 70+ hours of video recordings from phase 1.\footnote{Phase 1 consisted of 18 dyads, 1 triad, and 2 1:1s.}

\paragraph{Phase 2 – Digital Diary.} The second phase documented participants' daily routines and practices. Over one week, they each completed three exercises about how they engage with information: (1) how they found news/information on social media; (2) how they used search engines; and (3) how they related to information shared with them by others. Participants uploaded screenshots of social media and digital news content, and their own voice recordings and text responses to research prompts, describing their online information habits and views on information quality/credibility. 

\paragraph{Phase 3 – 1:1 Interviews.} In the third phase---conducted within one month of phase 1---we interviewed each participant one-on-one. We probed deeper into their information experiences, followed up on themes emerging from phases 1 and 2, allowed participants to revisit topics of their choice from phase 1
, and tested three information literacy intervention concepts.\footnote{Concepts were: accuracy prompts on a search engine, short information tips on TikTok, and longer videos explaining information literacy concepts.} Each session was conducted remotely and lasted 60 minutes. We collected 40+ hours of video recordings from phase 3. 

\paragraph{Incentives.} Each participant received \$425--\$500 USD as a thank you for their 5 hours of participation; parents/guardians of participating minors received \$75--\$100 USD. This reflects best practices from our market research partner.

\subsection{Analysis Methods} 

Our analytic approach used the principles and methods of constructivist grounded theory \cite{charmaz2006constructing, mills2006development}. Rather than testing for deficits or evaluating participants against preset frameworks, we adopted an emic (from-the-inside-out) lens \cite{scarduzio2017} to learn about what participants did with online information and why---from their perspectives and in their words. 

Preliminary analysis began immediately following phase 1. Researchers who observed the sessions took detailed notes, identifying insights and portions of the session that related to our RQs. During and after phases 1 and 2, we iteratively refined areas of inquiry and identified dataset gaps. This informed phase 3, allowing us to tailor 1:1 interviews to each participant, dig deeper into what dyadic interviews missed or occluded, and explore emerging hypotheses.

We coded insights from interview data using an iterative, inductive, and reflexive qualitative analysis method \cite{saldana2021coding}. We used descriptive coding in a series of thematic analysis coding passes, then tabulated each discrete code (\eg \textit{Admitting mistakes is a sign of trustworthiness}) against every participant whose interviews or diary submissions generated that code. In the codebook, we separated participants by age bucket as well as mainstream or alternative. Through collaborative discussion, we then performed several theoretical coding passes, iteratively clustering insights into 4 categories: social context, content, design and usability, and medium. We used the codebook to ensure data saturation, engage in rigorous collective qualitative analysis, and reflexively mitigate cognitive biases. We reached data saturation on the findings presented in this paper and suggest for future work to confirm reliability in Section \ref{sec-disc}.

Once data collection and coding were complete, we held several collaborative analysis sessions. Research team members wrote short memos to propose overarching arguments or explanations, and we iterated those interpretations through discussion. We referred regularly to the codebook to ensure that our analysis reflected the full dataset, draw conclusions about how insights applied to groups within our dataset, and identify new themes and meta-insights. 

\subsection{Research Ethics} 

When working with any population---especially when minors are involved---ethical considerations are essential. Though our organizations do not require IRB approval, we adhere to similar standards. Our study went through review to ensure that participants and parents of minor participants provided informed consent and were protected from undue risk. Our study plan was reviewed by experts in ethics, human subjects research, human subjects research with minors, legal, security, and privacy. Before finalizing our methods and instruments, we conducted an interdisciplinary literature review and consulted academics who specialized in engagement with information online, including by adolescents. 

When interviewing minors, two adults were always present. A parent/legal guardian completed consent forms on behalf of their child and provided verbal consent. We verbally explained the study to the minor and parent/guardian, confirming each understood. We reminded them that they could withdraw from the study at any time without consequences. 

We maintained strict data privacy controls for all participants. The market research agency assigned participants a pseudonym; all data collected (\eg notes, recordings, etc.) referred only to the pseudonyms. Participants were instructed to withhold personally identifying information (PII) during the study, and chose a pseudonym for the research team to use during interviews. Raw interview and diary data was scrubbed of PII during transcription. Raw research data will be deleted within 30 days of the final reporting and publication of results from this study. 

To mitigate the risk of participant identification, we have omitted personally identifying details, phrases, or words from quotes in this paper. Further, the pseudonyms used below are not the same pseudonyms used in data collection. 

\subsection{Limitations}

This study analyzes a small sample of experiences in-depth. Our sample is not statistically representative. During recruiting, we ensured that our sample did not significantly skew toward a particular demographic (except possibly mainstream/alternative viewpoints, the frequency of which should not be taken as representative). 

Known challenges with dyadic interviews include one person dominating the interview, pairs hesitating to discuss topics where they don't share values, and individuals self-censoring to preserve the relationship \cite{szulc2022practice}. To mitigate these, each participant completed phases 2 and 3 without their dyad partner. This study was also affected by standard limitations of self-reported data, including recall, observer, and social desirability biases. Pairing diary exercises with interviews should have helped mitigate such bias, since the multiple formats and phases enabled us to cross-reference and ask probing questions. Dyadic interviewing is known to reduce social desirability bias \cite{szulc2022practice}. A further strength is that the shared understanding between dyad partners from pre-existing relationships can lead to insights the interviewer might not otherwise be able to elicit \cite{christensen2004children}.

In some cases, the presence of others (\eg parent/guardian) may have affected responses. To account for this, we took notes that indicated the minor's awareness of a parent's or guardian's presence, capturing bodily and vocal cues, and any active participation from the parent/guardian \cite{christensen2004children}. This also helped us analyze the social effects of family influence on information-seeking practices. 

Finally, none of the authors are from Gen Z and therefore do not share the same subjective generationally-mediated experiences as participants. We mitigated this through weekly research discussions, including reflexive interpretations of our generational frames. We also consulted Gen Zers to help shape our study instruments.

\section{Results} \label{results}

In this section, we describe three key challenges Gen Z participants experienced when trying to make sense of online information: information overload, misrecognition, and social error. We then present the information sensibility practices they used in the face of these challenges, discussing the implications of and trust heuristics underlying each (Table~\ref{tab-overview}).

\subsection{Challenges}\label{sec-challenges}


\subsubsection{Information overload}

Participants felt deluged by facts, opinions, and ads as they encountered information every day. They wanted short, emotionally-manageable, personally-relevant, and uninterrupted experiences in information ecosystems that they could curate with little extra effort. Participants sought information in quickly absorbable forms. They also felt emotionally overloaded, attempting to control their information ecosystem to reduce encounters with information presented in a negative tone. This often meant curating an information ecosystem where they would only encounter such content. They also preferred to search for information \textit{within} that ecosystem when searching felt necessary. Many felt that they would naturally \textit{encounter} truly important or personally-relevant information, rather than needing to search for it. 

First, participants sought ways of finding information in short, digestible forms that could be easily and swiftly consumed. When encountering a long-form article in his diary exercise, Zak (18-24, Mainstream) noted: \iquote{It was too long to read, so I didn't even read all of it.} Leo (13--17, Mainstream) expressed a sense of information overabundance: \iquote{Sure, you can look on your own, but no one has time for that.} Many participants explained that search engines often ranked long-form articles highest, but those articles often felt like too much to absorb. Nayeli (18--24, Alternative) preferred to encounter and search for information on social media instead: \iquote{I like Twitter, just because it's quick to scan through information.} Many sought short-form content on social media, where they could swipe to the next video or scroll when overwhelmed or uninterested.  

This desire for speed cut across the video/text binary, contradicting the popular hypothesis that Gen Zers universally prefer video (\eg \cite{education2018beyond}). Several participants told us that they preferred a quick news article to a 10-minute video, because they could skim one but not the other. When encountering new information on websites or by creators they were not intimately familiar with, they preferred short text news-bites and short-form videos to longer video explanations or long-form articles. During a screenshare exercise, Sami (18--24, Mainstream) explained that she skims articles to \iquote{save time} because she \iquote{doesn't have time to read it all. I need to get down to the point.} She rarely clicks on longer videos: \iquote{I get bored really fast. I'm a very quick reader, and I don't have patience for video.}

\begin{table*}
    \centering
    \begin{tabular}{p{5.5cm}p{6cm}p{2.2cm}}
                                                                                                 \toprule
        \bf Challenges & \bf Practices & \bf Heuristics                                       \\ \midrule
        \textbf{Information Overload} & \textbf{Surrogate Thinking} & \textbf{Convenience} \\
        too many articles, facts, opinions, and ads, often too wordy, too long, and too negative & finding a trusted go-to source, and relying on the source's opinions\\ 
        \textit{\textcolor{royalazure}{---Period}} & \textit{\textcolor{royalazure}{---Period \& Age}} \\ \midrule
        
        \textbf{Misrecognition} & \textbf{Everyday Experience} & \textbf{Tone, Aesthetic}  \\
        lack of ability to speak to their generation, seen in tone, lack of controls, vibes & 
        privileging on-the-ground crowdsourced reporting, lived experience, first-person narratives \\ 
        \textit{\textcolor{royalazure}{---Age}} & \textit{\textcolor{royalazure}{---Cohort}} \\ \midrule
        
        \textbf{Social Error} & \textbf{Crowdsourcing Credibility} & \textbf{Tone, Aesthetic}  \\
        
        fear of making mistakes, taking unpopular positions, or actions that lead to social harm & 
        using comments, likes, and other peer discussions to evaluate content \\
        \textit{\textcolor{royalazure}{---Age}} & \textit{\textcolor{royalazure}{---Cohort \& Period}} \\[0.2em]
        
        & \textbf{Good Enough Reasoning} \\
        & limiting effort to finding just enough, a few facts, numbers, whatever is easy, and then moving on \\
        & \textit{\textcolor{royalazure}{---Period}}                    
        \\ \bottomrule
    \end{tabular}
        \vspace{0.5em}
    \caption{A summary of the three information challenges we found our Gen Z participants encountered, the four practices they used to navigate these challenges, the heuristics they applied, and, in \textcolor{royalazure}{\textit{blue italics}}, a summary of our hypotheses on the age, period, and cohort effects we discuss in Section~\ref{sec-effects}.}
    \Description[Summary of our Gen Z participants' information challenges, practices, and heuristics.]{Table 2 provides a summary of the 3 information challenges our Gen Z participants encountered, the 4 practices they used to navigate these challenges, the heuristics they applied, and a summary of our hypotheses on the age, period, and cohort effects we discuss in Section 5.3. The Information Overload challenge is summarized as too many articles, facts, opinion, and ads, often too wordy, too long, and too negative. We hypothesize that the period effect applies. Participants used the practice of Surrogate Thinking to navigate Information Overload. Surrogate Thinking is summarized as finding a trusted go-to source, and relying on their opinions. We hypothesize that both period and age effects apply. The heuristic they applied was Convenience. The Misrecognition challenge is summarized as the lack of ability to speak to their generation, seen in tone, lack of controls, and vibes. We hypothesize that the age effect applies. Participants used the practice of Everyday Experience to navigate Misrecognition. Everyday Experience is summarized as privileging on-the-ground crowdsourced reporting, lived experience, and first-person narratives. We hypothesize that the cohort effect applies. The heuristics they applied were Tone and Aesthetic. The Social Error challenge is summarized as the fear of making mistakes, taking unpopular positions, or actions that lead to social harm. We hypothesize that the age effect applies. Participants used two practices to navigate Social Error. One practice was Crowdsourcing Credibility which is summarized as using comments, likes, and other peer discussions to evaluate content. We hypothesize that both cohort and period effects apply. The other practice was Good Enough Reasoning which is summarized as limiting effort to finding just enough, a few facts, numbers, whatever is easy, and then moving on. We hypothesize that the period effect applies. The heuristics they applied were Tone and Aesthetic.”}
    \label{tab-overview}
\end{table*}

Second, participants often chose where they consumed information based on their assumed ability to manipulate algorithms and filter information sources present on their feeds. Using routine engagement to tailor their feeds was perceived to reduce the strain of sifting through information. They also believed that the information they encountered in those tailored information ecosystems was more likely to be personally interesting or meaningful. Tyse (13-17, Mainstream) trusted that if something was important, it would come across his feed. He explained that there was no need to search or follow news and politics, because \iquote{When stuff is important, it gets shared. When lots of people share, there has to be a reason.} Mo (13--17, Mainstream) said: \iquote{The more videos you 'like' of that topic, it will show you more, so you can decide what to interact with.}

This preference for algorithmically-tailored information sources reflected a desire for emotional management as well as relevance. Most participants expressed needing to periodically ``take a break'' from online information---particularly political information---because it negatively affected their mental health and well-being. Tyse (13--17, Mainstream) told us: \iquote{I just want to go online to have a good time […] I usually just scroll, and since TikTok personalizes your \#ForYou page, I just block [people who want to argue] because I don't really care to see that.} As we discuss further in \ref{sec-trust}, controlling the tone of information was critical for many participants.

Finally, encountering pop-ups felt disruptive, frustrating and overwhelming to many participants, so they avoided long-form news sites---including mainstream news sites. Describing her frustrations, Rosa (18--24, Alternative) told us: 
\begin{quote}``Clickbait is super annoying! Especially when I click on something hoping it'll tell me something, and then it's nothing related to the title. It's super misleading. They make those strong titles so you click on their page, you see the ads, they get paid for that. It seems like clickbait to get the views.''\end{quote}

Long-form news sites were often populated with pop-ups asking for donations, advertising, and GDPR (General Data Protection Regulation)/consent requests. As one participant put it, \iquote{Any ad is super-annoying when it takes my attention away. The ones that are distracting are definitely the most annoying.} Most felt social media ads were more seamlessly integrated into the user experience, rather than being an attention-grabbing imposition. Matt (13-17, Mainstream) said he liked a source if \iquote{it's not pushing you to sign up with an email to look at the article.} Like Rosa, ``clickbait'' demanding his attention also reduced credibility, as he continued: 
\begin{quote}``Anything that's even remotely clickbaity I ignore. It completely loses all credibility to me. Because to me, if you're making clickbait, you have zero faith in your content---you know, your content isn't actually clickworthy---so you're the bait. And news sources---even CNN and New York Times---do clickbait. I throw those articles away immediately.''\end{quote}

\subsubsection{Misrecognition}

Many participants felt misrecognized as a generation by major internet platforms and mainstream media. They disliked sources that did not ``speak their language,'' were aggressive in tone, or were not relevant to their interests. First, the aggressive, politicized language and content of mainstream media, and a lack of identification with presenters and pundits, pushed many to seek sources that ``speak their language.'' This often meant they avoided mainstream media in favor of more personally relevant content. As Mari (18-24, Mainstream) articulated: \iquote{I've seen the New York Times trending on Twitter for having silly things that don't matter. That makes me not want to visit the New York Times or USA Today.} Her dyad partner said she would watch the news once she was \iquote{really an adult,} but not now. 

Filtering for tone was a common response to misrecognition. Dina (18--24, Mainstream) explained that Gen Zers feel alienated from mainstream news sources due to the political partisanship, punditry, and agendas, calling it \iquote{noisy.} As she explained, \iquote{I want to know Russia invaded Ukraine, not whether it's the Democrats' fault or the Republicans' fault.}

Participants noted that they could not control the tone of results on search engines, but they could manipulate their social media algorithm, lowering the likelihood that ``noisy'' sources would appear in their feed. As Nella (18-24, Mainstream) put it: 
\begin{quote}``Currently where I'm located on the algorithm, it's pretty light-hearted. So to try and beat it, I'd end up in a spot where I'd be extremely uncomfortable and be surrounded with things I'd rather not interact with… [The creators she follows on TikTok] know how to balance the bad news with the good news. I find if I'm too much exposed to bad news, I get down and feel really drained. I just don't take in as much information when I'm feeling drained. I scroll past. Anything that has to do with real life, I will 100\% scroll past.''\end{quote}

Second, the tone and style of many information literacy interventions (\eg videos explaining information literacy concepts, automated fact-checks, etc.) felt patronizing to participants, reinforcing their belief that they did not need help identifying misinformation. As young people on the cusp of adulthood, most participants felt completely misrecognized by infantilizing and/or pedantic language, images, style, and music. This belief was exemplified by a participant who reacted to one information literacy video by saying that it \iquote{isn't for me. This is for someone who doesn't understand information. Who doesn't have basic information literacy? Who doesn't know how to search for things?}

Many thought grandparents or younger children would be good recipients for information literacy training, seeing it useful only as a tool to teach those who were less informed. As Zak (13--17, Mainstream) asserted, \iquote{Gen Z knows all of the "invisible rules" of the internet.} Even if technology companies had something to teach his peers, he said, they \iquote{haven't found the right way.} He disparaged animated video interventions by saying: \iquote{My generation sees this as an attempt to relate---[but] if you're making an attempt to relate, you're fundamentally unrelatable.} 

Finally, news and information found via social media or discussion boards was appealing to participants because it was accompanied by an apparatus for gauging social context, meaning, and significance. We discuss this next. 

\subsubsection{Social error}

Finally, participants perceived a potentially steep social cost of being wrong---social error---leading to risk-aversion. To cope with their fear of social error, they checked comments for social orientation and searched for answers validated by peers. They often remained anonymous online to avoid being ``canceled.'' 

The fear of not belonging drove them to check comments to orient themselves socially. They worried about sounding misinformed, so they searched for signs that answers were not only ``true,'' but had been accepted and validated by their peers. They found it difficult to find these peer validations using search engines and traditional news sites (many of which have limited or removed comment functions and/or lack detailed profile information attached to the commenter) but easier on discussion forums and social media. As Matt (13--17, Mainstream) put it, \iquote{if you're out of the loop, you're out of the group.}

The threat of being ``canceled'' caused some participants to hide posts from friends and family, instead staying anonymous online. The stakes of ``being right'' in social terms when assessing and sharing information felt extremely high. This threat was poignantly exemplified by Kyle (13--17, Alternative) who had posted content that he was pretty sure \iquote{people online liked.} Soon, however, he discovered that not everyone at his school agreed. After being doxxed then intimidated on the street by strangers, he transferred schools. Still in his early teens, he wanted to \iquote{get back to the earlier days} when things were simpler. Expressing a similar sentiment, Dani (18--24, Alternative) said: \iquote{I don't post, because it's horrifying to post something and then get 30 people telling you why you're wrong.}

The clearest example of a positively-altered information practice that we heard from participants occurred through a social intervention that responded to a participant's need for social belonging, not for higher-quality information. Kelly (18--24, Alternative) told us how they started getting into a Neo-Nazi movement. They had multiple people along the way online try to argue with them, debunk their reasoning---but they ignored all of it. If anything, it fueled their sense of opposition. The only friend they kept in touch with, they explained, did not try to dissuade them. Instead, she invited them to play Dungeons \& Dragons (D\&D). It was this social experience that pulled them out of Neo-Nazi rabbit holes, they explained, because they got really into the D\&D community. The D\&D community replaced the sense of community that they had been looking for from the other movements. They also said that D\&D helped them learn empathy because it was a role-playing exercise.

Next, we share the practices and trust heuristics participants used to navigate the challenges we detailed above. 

\subsection{Practices \& Trust Heuristics}

Participants used specific practices and trust heuristics to overcome the challenges detailed in Section \ref{sec-challenges}. We argue that Gen Zers interpret information as aspirational members of social groups, not as isolated individuals. They use and seek online information to help navigate their social environments, not just to establish truth or find facts. Here, we present 4 practices---crowdsourcing credibility, surrogate thinking, "good enough" reasoning, and everyday experience---and describe how Gen Zers rely on the trust heuristics of convenience, aesthetic, and tone while using those practices. 

\subsubsection{Crowdsourcing credibility}

Participants crowdsourced their credibility judgments\footnote{The term "crowdsourcing credibility" was used by Giudice \cite{giudice2010crowd} to describe user interactions with web pages that visualize crowd evaluations of the credibility of news content. We offer an extended version of this concept here, finding it to be an ecologically-valid practice that is actively and dialogically undertaken by users across a range of platforms and online spaces, not simply in response to particular affordances.} by observing how others reacted to the same information. By reading comments or tallying likes on a post, they used the judgment of their peers and a sense of ``the discourse'' to socially orient themselves and decide what they should think. Beyond truth-seeking, they wanted to know what beliefs were normative around them. Evan (13-17, Mainstream) explained:
\begin{quote}``It's important to get the bigger picture, because you might want to confirm that what you're thinking is the widely accepted opinion, or maybe you want to go into the comments to see if other people are thinking differently…that could give you a better view.''\end{quote}

The participants liked to take the social temperature of how users ``like them'' responded to content. As they encountered information, they contextualized their interpretations by looking for markers of social similarity and validation. They asked: What do people like me think about this? As Shelly (13--17, Mainstream) articulated: 
\begin{quote}``Most of the time, I use the comments to see what other people are saying, if they're confirming it or denying it or correcting it. I try to double-check my sources with the comments, by Googling, and with other content creators and what they say.''\end{quote}

Note that her practice is to go to the comments first and then Google to confirm what she finds there, using a search engine as a supplement to multiple social checks---a behavior that has also been seen in adult users while fact-checking~\cite{geeng2020fake}. Shelly then checks with her surrogate thinkers (a practice described in \ref{sec-surrogate}): content creators. We observed this practice frequently. Many participants heavily weighted comment sections alongside formal web page content, dialogically reading between them. This "back-and-forth" between comments and content illustrates how Gen Zers use a form of lateral reading \cite{wineburg2022lateral, breakstone2021lateral, caulfield2017web} to pursue the social aims of information sensibility rather than information literacy. 

Max (18--24, Alternative) also illustrated her trust in this comment-checking practice--as well as her potential susceptibility to misinformation. She explained: \iquote{I go into the comments just to verify if it's true or not, because usually if I go into the comments and something's fake, somebody knows about it.} As Adam (18--24, Mainstream) put it, \iquote{If people in comments are saying it a lot, it gets more validated. If all these people have the same info, it had to come from somewhere.}

\textit{Likes} were another means of crowdsourcing credibility. As Isabel (18--24, Mainstream) explained: 
\begin{quote}``This is bad to say, but if the video has more likes, then I'll probably believe it more. If it's just a random person who showed up on my feed and they have no likes, then it's like "who even is this person?"''\end{quote}

For participants, the most useful searches were shared experiences, when they could connect with others (in-person, virtually, or asynchronously) while seeking and evaluating information. Searching ``does water get dusty?'' during a screenshare exercise (a query chosen by a participant), one 13--17 Mainstream dyad told us they love Google's \textit{People also ask} feature. They explained that sometimes other people's wording is better, that others' questions can be ``\textit{super smart}.'' The dyad appreciated that this feature resembled the crowdsourcing they already practice.

\subsubsection{Everyday experience}

Participants valued everyday experience when evaluating information online. As they felt traditional gatekeepers and authorities losing trust and social traction, participants used markers of "authentic" experience to judge credibility over markers of expertise. Participants looked for evidence that a source was ``on the ground'' or had relevant personal experience. They valued what people could see, hear, and feel for themselves.

\begin{figure}
    \centering
    \includegraphics[width=0.47\textwidth]{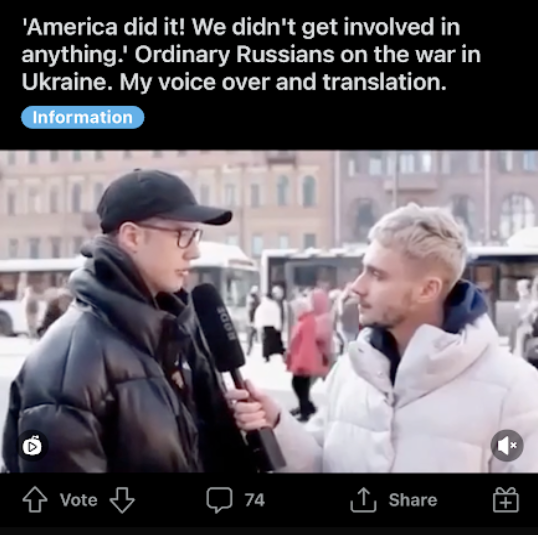}
    \caption{A source from Reddit where an unknown reporter appeared to be interviewing people on the street in Russia, increasing credibility from the perspective of the participant who shared this with us.}
    \label{fig-ukraine}
    \Description[A participant's example of a "credible" source from Reddit]{Figure 1 is a screenshot---probably taken on a mobile phone---of a source from Reddit where an unknown reporter is shown interviewing a person on the street in Russia. The interviewer appears to be a young adult man wearing a puffer coat. He is holding a microphone up to the mouth of another person who appears to be another young adult man who is wearing a puffer coat and has on a baseball hat and glasses. Behind the two men and slightly out of focus are various pedestrians and a row of buildings. This on-the-street image increased credibility from the perspective of the participant who shared this with us.}
\end{figure}

One of our diary exercises illustrates this valuing of everyday experience. Ari (13--17, Mainstream) was eager to learn what Russians thought about the invasion of Ukraine. He shared a Reddit source (Figure~\ref{fig-ukraine}) as an example of a source he deemed to be reliable. He wasn't concerned that he didn't know the reporter's name, background, or credentials. It sufficed that the reporter appeared to be, in his words, \iquote{on the ground ... [talking to] everyday Russians.}

Personal accounts felt more genuine and relevant to participants' everyday lives. As Sami (18-24, Mainstream) articulated: \iquote{People of my generation and younger prefer TikTok and real people talking... It's more personal [and] feels more genuine.}

One 13--17 Mainstream dyad told us when investigating the rumor that Katy Perry killed a nun (screensharing with us a query they chose), they were disappointed to find no stories from major news sources that definitively answered this question. They went to TikTok and concluded that if Katy Perry fans hadn't weighed in, the story must not be true. They trusted Katy Perry fans, who engaged with and reported on her activities daily, to know the truth. A lack of information on trending topics on search engines led several participants to turn to social sources of information. 

\subsubsection{Surrogate thinking} \label{sec-surrogate}

Many participants, especially the alternative media users, had a ``go-to source'' that they felt they could trust: a person with similar values who could filter information and clarify points of view. Gen Zers trusted that this person had done the work of researching and thinking for them. This helped offset the uncertainty participants felt about institutional agendas (\eg media, corporate, partisan, etc.) while also providing them with a sense of social community and belonging. We term this practice \textit{surrogate thinking}. Often these surrogates were those with everyday experience, as discussed earlier; others were selected because they helped protect against social error.  

Participants used go-to public personalities and influencers as surrogate thinkers to help them filter and interpret information. As Ryan (13--17, Mainstream) summarized: \iquote{In our generation, [people] tend to align with information because they align with the people giving it to them.} This was a direct response to the challenge of information overload: many participants preferred a go-to source because they were overwhelmed by the volume of online source material, felt generalized skepticism, and were uncertain about unfamiliar sources. One participant (18-24, Alternative), for whom Jordan Peterson was a go-to source, said: 
\begin{quote}``You know, I generally take [Jordan Peterson] at his word. I know he's intelligent and well-read, but I guess [after participating in this study] I'd be more inclined to fact-check than in the past. I've just taken him at his word because I know he's very well-read.''\end{quote}

We heard this sentiment expressed about a range of surrogate thinkers. Joe Rogan came up repeatedly: 
\begin{quote}``I'll usually watch Joe Rogan to get more information. I can tell when he's genuine because he does bluff on himself. You can tell when he's lying. He will also call himself out to say that he's wrong---he'll say that, he'll admit it.''\end{quote}

Surrogate thinkers were not necessarily famous online figures or influencers; they were often family members. Some participants heavily weighted the opinions of authority figures in their own family while gathering information and developing their opinions. Family members often played a significant role in helping participants triage information and judge the appeal or credibility of other media sources. As Bella (18--24, Alternative) articulated:
\begin{quote}``[My uncle] likes being up-to-date about what's going on, so I ask him, and he sends me the videos that he found, the articles that he found. A lot of my messages with [him] are just back and forth articles.''\end{quote}

Go-to sources themselves were often trusted based on input from family members. These shaped the searching practices of participants, limiting where they might look to validate information. As Cris (18--24, Alternative) explained, speaking about a source shared by a family member: \iquote{I go to ZeroHedge because they basically have everything I'm looking for. If I go to Google, usually I'll type ZeroHedge into the search box too.}

Many participants emphasized that the news consumed by older family members shaped their own sense of what was credible online, as well as what might be socially acceptable or valuable. Whether watching Newsmax with grandparents or listening to NPR with parents before school, experiences in the family home with news media shaped how participants judged information online. As Melissa (18--24, Mainstream) explained, expressing a position we heard from many participants: \iquote{When my mom has an opinion, then I'll probably agree with her […] that's just how I've been raised.} Over time, several participants traveled away from family influences when they encountered a strong enough "pull" from another social influence (\eg new friends, college professors, etc.).  

In these ways, selecting a surrogate thinker addressed participants' fears of social error, because the go-to source aligned with their values and could help them find what positions might be socially acceptable. Participants practicing surrogate thinking who used family members as go-to sources experienced a stronger pull toward the orbit of their immediate social circle. Even when this worldview was shaped by misinformation, they were reluctant to reject it for fear of social error.

\subsubsection{``Good enough'' reasoning}

Participants used a practice we term \textit{``good enough" reasoning} to calibrate the effort necessary to find and verify information. This calibration depended on the social purpose and meaning of the information (what they wanted to do with it) and the degree to which participants were willing to tolerate uncertainty. Few participant information journeys ended with a definitive ``true'' or ``false'' answer. The intended social use of the information mattered. For participants, having ``good enough'' information meant knowing enough to win an argument (\eg with a parent or on Twitter), learning enough about a topic to feel confident talking with friends about it, or understanding enough to avoid making a mistake or sounding out-of-the-loop. 

Alexa (18--24, Mainstream) spoke at length about her process of information-seeking, explaining that she mostly searched for news to win arguments with her father. Her politics started to shift when she moved away from home, and she reported getting into arguments via text message with her father about a range of topics. She explained that when she actively uses a search engine for content, it's always in response to a specific article her father sends her (\iquote{usually his articles are blatantly biased; they're full of insults like "sleepy Joe Biden"}), and she's searching for a quick, unbiased response (\iquote{something fact-based, a number-type thing}) that can end the argument. She stops searching when she finds a fact that she deems is "good enough" to share with her father. 

Social utility shaped the importance and purpose of information-seeking for participants. As Shantal (18--24, Mainstream) said when discussing the importance of having the ``right'' information:
\begin{quote}``Back up what you say. Make sure you don't look like an idiot on the internet spewing something, then it turns out you're completely wrong. You could have a very non-factual take on something that could affect the way people view you […] But it's also OK to just not know or care if it's not relevant to you.''\end{quote}

The use of information to signify social belonging and perform impression management (not ``looking like an idiot'' or ``affecting the way people view you'') is clearly demonstrated. Further, getting accurate or factual information often only mattered to participants if it was ``relevant'' to their specific interests. For Mo (13-17, Mainstream), information was important if \iquote{it has an effect on me, or something that is going on around me […] you make your own call on how important it is to you.}

Importantly, participants had different stopping-points on information journeys based on their individual comfort with uncertainty, since many of their journeys did not quickly yield definitive answers. The more comfort they felt dwelling in that uncertainty, the more likely they were to conclude a search with a decision that not knowing for sure was ``good enough.'' 

\subsubsection{Trust heuristics}\label{sec-trust}

In this section, we describe the trust heuristics participants used to determine whether sources were credible, useful, and/or meaningful. Trust heuristics are sets of informal criteria, or shortcuts, that people use to assess information. These shortcuts---like folk models \cite{wash2010folk, d2005some}---do not necessarily lead to accurate information, and they can even make misinformation seem compelling. Conversely, they can also help expedite decision-making and serve important social functions. Participants used three key trust heuristics: convenience, aesthetic, and tone. 

\paragraph{Convenience} The degree to which the information is easy to access and share. Convenience is a response to information overload, as Maya (13--17, Mainstream) said: \iquote{Searching on Google is a hassle, it's more for school. TikTok is more fun. The videos are more interesting and relatable […] On TikTok I don't have to read mounds of paragraphs.}

\paragraph{Aesthetic} The degree to which the information ``feels right.'' Pop-up ads were a key negative aesthetic, as Clio (13-17, Mainstream) articulated: \iquote{When there are a lot of ads, you can tell that they just care about making money and not the actual news.}

\paragraph{Tone} The degree to which the information is affectively charged or not -- especially negatively. As Shay (18-24, Mainstream) explained: \iquote{I think the news is negative. Sometimes I tell [my grandfather] that and he doesn't understand.}

\begin{figure}
    \centering
    \includegraphics[width=0.42\textwidth]{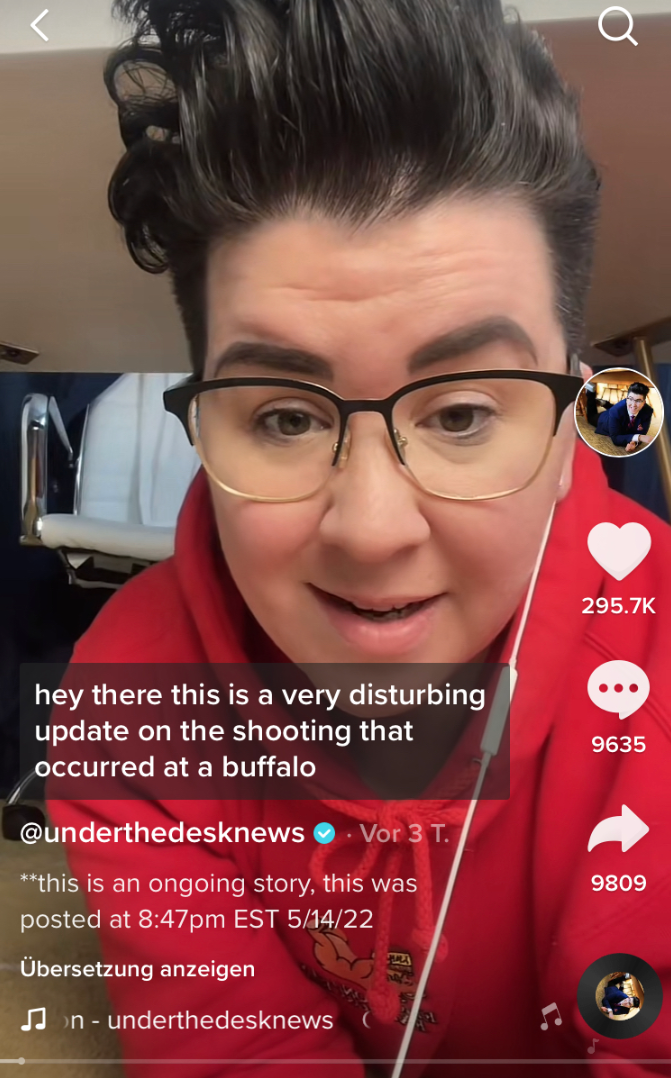}
    \caption{Example from @underthedesknews on TikTok providing calm, quietly-spoken news from underneath a desk, about a recent shooting in Buffalo.}
    \label{fig-underthedesk}
    \Description[A participant's example of a good news source from TikTok]{Figure 2 is a screenshot---probably taken on a mobile phone---of an example from @underthedesknews on TikTok. The image is of a close up of a young adult who is wearing glasses and a hoodie. They appear to be under a desk with a desk chair partially in view behind them. The text on the image says “hey there this is a very disturbing update on the shooting that occurred at a buffalo” then some additional text about the post. A heart icon shows 295,700 likes, another icon shows that 9,635 users have commented on the post, and 9,809 have reshared the post. The participant provided this as an example of calm, quietly-spoken news about a disturbing current event.}
\end{figure}

Multiple participants highlighted @underthedesknews on TikTok (shown in Figure~\ref{fig-underthedesk}) as a trusted news source; it showcases all three heuristics. Convenience: it gives quick and snappy overviews. Aesthetic: young, informal, relatable individuals report the news in place of "the usual" talking heads on mainstream media. Tone: it is gentle and reassuring, making upsetting news easier to handle. As Jax (18--24, Mainstream) described: \iquote{It's not condescending. It's bringing you the day's news in a calm way from a safe space.}

\section{Discussion}\label{sec-disc}

The central finding in this work is that online information processing is fundamentally a social practice for Gen Zers. Therefore to be effective, information literacy interventions should treat individuals as aspirational members of social groups. While other works have noted the role of social influence on credibility judgments, we argue that it is not simply that social ties affect users when they evaluate online information---social belonging itself is often the primary motivation for information engagement. In this section, we relate these findings to our original RQs as well as prior work, and we consider implications for intervention. We also hypothesize which phenomena are period, age, and cohort effects, suggesting where our findings have more general implications.

\subsection{Implications of Information Sensibility}

Here, we discuss three implications of our work that address our first and second RQs (how Gen Zers seek and assess online information): (1) Gen Zers' informational and social needs are inseparably entangled; (2) Gen Zers’ information journeys often do not begin with a truth-seeking search query; and (3) Gen Zers use information to orient themselves socially and define their emerging identities. We handle RQ3 (susceptibility to misinformation) in \ref{sec-misinfo}.

\paragraph{Gen Zers’ informational and social needs are inseparably entangled.} 

Other works have called attention to the importance of social influences on news consumption~\cite{messing2014selective, giudice2010crowd, qayyum2010investigating}. We take this work on social influence a step further, arguing that it is difficult to disentangle the process of searching for facts from the process of seeking social support. Much existing literature assumes a clear distinction between online truth-seeking (looking for factual information) and online support-seeking (using the internet to find community). Yet as Patel et al. have shown~\cite{patel2019feel}, such a distinction often does not hold. Like the participants in that work---who came online to make sense of their medical diagnoses, simultaneously seeking health information and social support or belonging---Gen Zers practice \textit{information sensibility}, personalizing and socially-interpreting ostensibly factual information to process it meaningfully. Work on communication ecology~\cite{wenzel2019verify, ball2012understanding} using communication infrastructure theory 
~\cite{ball2001storytelling, kim2006civic} instructs a similar attention to users’ goals and information pathways as they engage resources within multimodal environments to meet their needs.

We therefore call for further attention to questions of motivation and meaning when analyzing young people’s susceptibility to misinformation: why (in addition or adjacent to validating truth claims) young people encounter, seek, and metabolize information online; and in what contexts or circumstances information becomes meaningful and relevant for them. For Gen Zers, navigating online information is challenging because they seek social belonging as much as they seek truth, in many cases not marking a clear distinction. Our participants mostly valued information based on its relevance to their social relationships. As they encountered information, they assessed it in relation to their peers, family, and go-to sources, interpreting (or disregarding) information based on the pull of these social influences. 

Citing the classroom as a key influence, boyd~\cite{boyd2018you} has argued that “perverted” media literacy, as students learn it in school, teaches students to “question everything” without filling resulting gaps with new frameworks to make sense of the world. We observed similar “perversions” of media literacy principles in our participants’ interpretations: the belief that Wikipedia is universally unreliable, the push to “question everything” and “consult a wide range of sources” leading our participants to fringe (mis)information sources, and the belief that many mainstream media sources have hidden agendas. boyd argues that misinformation peddlers online weaponize critical thinking and easily manipulate the gap created by “questioning everything,” leaving youth susceptible to misinformation. 

Yet we found that rather than school being a uniquely destabilizing moment, where students learn critical thinking which unseats their prior beliefs and leaves a vacuum, young people are constantly being influenced by their social ties. They dialogically use information to make sense of their social place in the world (and vice versa), making their truth-seeking processes largely inseparable from their social validation-seeking process. Lifestage changes that bring new social influences (\eg going to college, starting a new job, entering a romantic relationship) appear to be moments when reorientation of social and informational needs is possible. These are often gradual shifts, as new social motivations may begin to draw Gen Zers into other interpretive frames.  

Our findings about the entangled nature of Gen Zers’ informational and social needs lead us to value solutions less strictly focused on classroom-based learning environments. Rather than resolving cognitive glitches and encouraging critical distance, as boyd argues, we suggest that social support and belonging may be key solutions to the problems of misinformation and fake news. Our participants were largely less focused on trying to fill knowledge gaps (created, as boyd proposes, by literacy education's drive to produce critical thinkers) than they were on social gaps, which they sought to fill by consuming the correct, most relevant information. A lack of social belonging pushed participants to seek new social groups, which they could access through shared information consumption. For many Gen Zers, in other words, the core problem is not misinformation (solved via literacy-based competencies) but rather social belonging, which may be solved via sharing misinformation.


Gen Zers make judgments about how seriously to take online information---and therefore how to evaluate it--- depending on its significance for their lives. Our work also shows that the evaluation of information, in situ, is not reducible to questions of facticity or accuracy for Gen Zers, but is tangled with judgments of relevancy and social significance. Prior work has shown that relevance, defined as being related to one’s interests, matters for misinformation behaviors among teenagers on messaging apps~\cite{herrero2020teens}. A wider approach to relevancy that incorporates social pulls could help HCI researchers better gauge not only whether Gen Zers are practicing information literacy, but when and why they choose to do so (or not).
The entangled nature of social and informational work for Gen Zers, as young people make sense of information together, means that HCI and literacy-based interventions should look for ways of supporting youth in their social or relational needs through engagement with online information, rather than presuming a primary need for help with truth-assessment.
\paragraph{Information journeys often do not begin with a truth-seeking search query.} Although we began this research interested in information-seeking behaviors, we found that Gen Zers largely 
\textit{encounter} information online, rather than actively seek it out. As one 13--17 mainstream participant put it: 
\begin{quote}``I use social media to keep up with current events, because it'll come across my feed. I don't sit down and watch the news every morning. I'll just use TikTok or Instagram to see what's happening.''\end{quote}

We found that many researchers analyzing information literacy implicitly use the concept of an ``information journey'' as a heuristic: a linear process that begins with a verifiable question entered into a search engine, where the user finds relevant information or content, performs a credibility check, and arrives at an answer (a model owing to the concept's origins in library and information science~\cite{american1989presidential}). The user moves from a state of being less informed to more informed (or incorrectly informed, in the case of misinformation discovery).

\begin{figure*}
    \centering
    \includegraphics[width=0.95\textwidth]{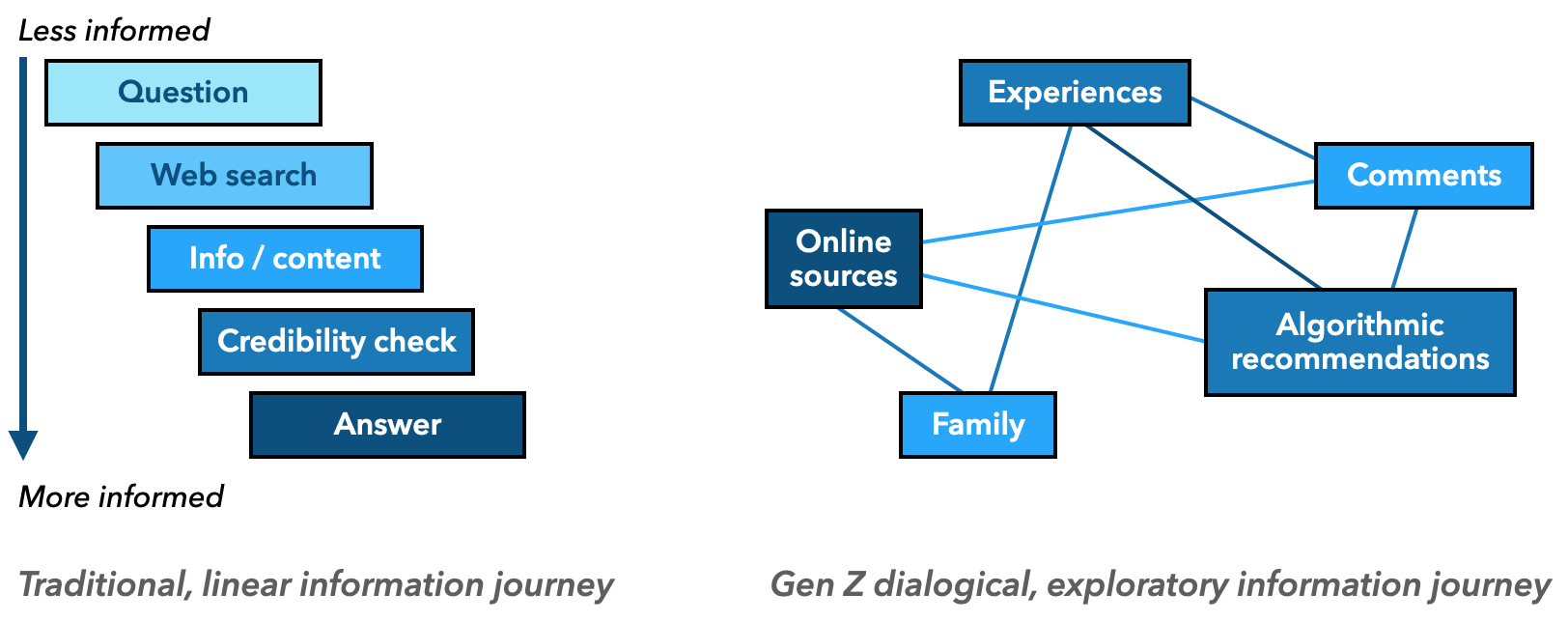}
    \caption{Two representations of information journeys. Left: A traditional, linear journey from question to answer. Right: A more exploratory and less definitive journey that toggles between information-seeking and information-encountering.}
    \label{fig-journey}
    \Description[Two representations of information journeys.]{Figure 3 shows two representations of information journeys. On the left, a traditional, linear journey goes from Question to Web Search to Info / Content to Credibility Check to Answer along an axis of less informed (at question) to more informed (at answer). On the right, a Gen Z dialogical, exploratory information journey is shown via 5 interconnected boxes. The boxes are Experiences, Comments, Algorithmic recommendations, family, and online sources. Each box connects to two or three other boxes to show how Gen Z toggles between information seeking and information encountering.”}
\end{figure*}

We discovered, however, that our Gen Z participants more often encountered information “in the wild” (rather than starting with a query and following a single search pathway), and often toggled dialogically between implicit information-encountering and explicit information-seeking, starting and finishing their engagements during conversations with family, in comments sections, on social media feeds, and elsewhere. We represent this nonlinear model in Figure~\ref{fig-journey}.

Prior work has identified this experience of passive exposure to information brought about by the internet; Qayyum et al.~\cite{qayyum2010investigating}, for instance, found young people ``encountering'' the news while surfing web portals (\eg MSN, Hotmail), more than actively seeking online information. Our work with Gen Zers finds that this practice has grown more significant with increased mobile and social media usage (especially the development of algorithmically-determined scrolling feeds, both text and short-form video) and should be incorporated into HCI and literacy-based information models as the default mode of engagement, not the exception to the rule. 
 
Furthermore, ours is not the first work identifying nonlinear information-seeking behavior; other studies, in both HCI and information science, have documented it empirically (\eg cyclical patterns in health information-seeking,~\cite{patel2019feel}) or sought to model it~\cite{foster2004nonlinear}. However, it remains common among studies of information literacy behaviors, especially among young people, to evaluate competencies premised on (and designed to support) more linear models of searching and evaluation (\eg~\cite{breakstone2022civic, hargittai2012searching, mcgrew2018can}). Ecologically-valid interventions that seek to channel or enhance nonlinear online information habits rather than eliminate them (\eg lateral reading~\cite{breakstone2021lateral,wineburg2022lateral}), could be more sustainable and effective among Gen Zers than approaches designed for linear journeys.

\paragraph{Gen Zers use information to orient themselves socially and define their emerging identities.} We investigated not just how Gen Z seeks and evaluates information, but also its function. We found that Gen Zers use online information not simply to verify facts, but for identity-formation and to achieve social belonging. In other words, information helps them determine who they are, where they fit, and what they believe. 

The use of online information for social orientation and identity-formation manifested in several ways among our participants. For many, finding and consuming the ``right'' information was a way of participating in social groups or building relationships with family members. Crowdsourcing the comments to analyze credibility was one key way to seek that social validation. For others, online information helped them push away from existing relationships (\eg with parents) by giving them tools to win arguments. We also found a widespread sense that engaging with online information relating to politics or the news was a signal of adulthood and maturity---whether already achieved or an aspirational future state. Many proudly stated that they were the ``kind of person'' who knew to check sources and evaluate credibility.

The literature on social semiotics, self-presentation, and identity-construction is vast, but we follow Taylor~\cite{taylor1989sources} in linking identity to practices of belief-formation. In keeping with social semiotic theories, we found that Gen Zers figure out what they believe not only by seeking and then internally processing facts about the world, but by dialogically orienting themselves with and through a much wider field of socially-meaningful, multimodal information \cite{bakhtin2010dialogic}. Learning from our participants to adopt a more capacious definition of information (both textual and audiovisual content, including facts, opinions, arguments, and social signals) allowed us to more fully understand this social semiotic process. This suggests the potential dividends for HCI and literacy-based approaches in adopting a richer and more inclusive definition of online information. 

These findings are also an empirical contribution to the HCI literature on identity and information. What we found with Gen Zers is a generalized form of what Tripodi~\cite{tripodi2020searching, tripodi2022propagandists} has found specifically for American evangelical conservatives: that online information practices may be used to signal political or social identities. Crowdsourcing comments for markers of social similarity is a key example of this practice. Further HCI studies might explore this for other cohorts (\eg older users or progressives) or examine how platform-specific features and affordances could shape identity signalling via information-processing.

\subsection{Susceptibility to misinformation}\label{sec-misinfo}

Here, we consider how this information sensibility framework accounts for the susceptibility of Gen Zers to misinformation (RQ3) and might reorient HCI and literacy-based approaches to intervention. We found that all Gen Z participants, whether alternative or mainstream, used information for social belonging. Three information sensibility practices we identified (crowdsourcing credibility, everyday experience, and surrogate thinking) were not strongly linked with either alternative or mainstream information habits but were effectively neutral; they often reinforced beliefs normative in participants' social circles, whether based on facts or misinformation. We posit that one practice (``good enough'' reasoning) seems to offer a protective effect for both mainstream and alternative Gen Zers, in cases where no definitive ``right'' answer was available. 

We began this work assuming that belief in misinformation would be non-normative for Gen Zers. Yet some Gen Zers exist in social or physical spaces where misinformation is the norm. Here, too, their information practices were affected by the desire to belong. We observed Gen Zers sharing misinformation when it was normative within their communities to do so. If most people around them were skeptical about COVID vaccines, for instance, they were more likely to share vaccine misinformation, out of a desire to connect with others. This finding---that Gen Zers' susceptibility to misinformation is deeply shaped by family influences or social contexts---extends previous work on the social appeal of misinformation~\cite{,messing2014selective,geeng2020fake,oconnor2019misinformation,duffy2020too} by clarifying the social influences at play on young people in particular (like~\cite{herrero2020teens}). It also confirms the need for sociotechnical approaches to misinformation in which interactional or pedagogical interventions are combined with attention to social forces, both on- and offline. 

In alternative information contexts, the practice of surrogate thinking frequently led Gen Zers to accept and share misinformation. Those whose surrogate thinkers were family members, like Bella (18--24, Alternative) and Cris (18--24, Alternative), quoted in \ref{sec-surrogate}, were urged to share vaccine misinformation or trust alternative news sources advancing conspiracy theories. Yet surrogate thinking can also lead to surprising shifts in orientation. For example, Hannah (13--17, Alternative) had a religious family that shared misinformation about COVID and Black Lives Matter (BLM). Hannah was also a big Justin Bieber fan, partly since he openly spoke about being religious. Because of this fandom, she listened closely to what Bieber said about other topics, like COVID. Since she ``trusts what he says'' about faith, she also trusted his claim that COVID was real, prompting her to rethink inherited family attitudes. As an information practice for Gen Zers, surrogate thinking confirms prior work on social signaling and trust~\cite{tripodi2022propagandists,vanderdoes2022trust} which has found, for instance, that how speakers make claims or signal group memberships  (\eg faith or ideology), matters as much, if not more, for trust than the substance of a given claim.

Crowdsourcing credibility, used by both alternative and mainstream Gen Zers, is another practice that involves making sense of social signals beyond the factualness of information. It is also a case---as others have found (\eg~\cite{tripodi2020searching})---of information literacy practices being deployed in novel ways that can increase susceptibility to misinformation. Participants often sought to verify facts in the comments (\eg by following links, or perhaps more typical, simply noting that links were provided), but not whether or not a commenter was who they said they were (or was even a real person). They knew about “fact-checking” news sources, but generally did not apply the same reasoning to individual influencers or commenters, rather evaluating signals of social similarity. Valuing the comments so highly could make Gen Zers of all political orientations more susceptible to misinformation, as this information space is deeply shaped by bots, bad actors, and incentives that could drive misinformation-sharing (\eg audience monetization, or a welcoming in-group). The Gen Zers in our study did not demonstrate a robust awareness of these risks, nor did they have sophisticated heuristics for judging the authenticity or motivations of accounts posting comments. (Valuing everyday experience, in this context, can also increase susceptibility to misinformation, if not paired with critical attention to authenticity.) By widening our definition of ``meaningful'' information to include social signals and collective beliefs, found often by Gen Zers in the comments, our work suggests how literacy-based interventions could be reworked with information sensibility. For instance, Gen Zers might benefit from accuracy prompts (found effective by, \eg~\cite{pennycook2021shifting}) within comments sections, or advice for considering the authenticity of digital identities.

We found that ``good enough'' reasoning may offer both alternative and mainstream Gen Zers a layer of protection against misinformation---especially when queries are not easily or definitively resolvable. For example, Alia (18--24, Mainstream) was a Reserve Officers' Training Corps (ROTC) member who told us about her process of figuring out whether COVID-19 had been made in a lab: her ``right-wing'' parents thought it had been, but her ``left-wing'' cousins found this claim to be ridiculous. She spoke to a friend with similar ``moderate'' beliefs as her. They searched online together but could not come to a definitive answer. Weighing all that they had found, they concluded that COVID-19 was probably not manufactured in a lab. Later that week, Alia attended an ROTC training exercise that simulated the release of a manufactured bioweapon. This made her question her initial belief: \iquote{Oh man, if the military thinks this is possible, then maybe COVID was manufactured in a lab.} Instead of following a misinformation rabbit hole, however, Alia concluded that she couldn't be sure. Her willingness to declare an information journey ``good enough'' meant she did not turn to misinformation for a definitive answer. 

``Good enough'' reasoning allows Gen Zers to stop information-seeking journeys before they reach a definitive answer to their query when they decide they have sufficient information or that certainty cannot be achieved (\eg for interpretive, rapidly-changing, or normative topics). This runs against the grain of literacy-based approaches, which aim to help people confidently complete linear information-seeking journeys by locating accurate, high-quality information. Yet many studies have identified a relationship between desire for certainty (termed by psychologists as the need for cognitive closure) and susceptibility to misinformation or conspiracy thinking~\cite{ecker2022psychological, marchlewska2018addicted, douglas2017psychology}. In many cases, misinformation preys on this dynamic by offering the false promise of certainty. Our findings suggest the need for interventions that build comfort with uncertainty or enable ``good enough'' reasoning, rather than aiming to reduce it on the basis that a single, discoverable truth always exists. 

Indeed, the Gen Z information sensibility practice of valuing everyday experience also illuminates the need to help Gen Zers navigate the information landscape as they find it, not as we might wish it to be. As Doctorow \cite{doctorow2017propaganda} argues, the misinformation crisis is not simply about what is true, but how we know what is true. Experience, as~\cite{boyd2018you} has written, is an epistemology recognized by disability activists and some Indigenous communities as a privileged mode of knowledge construction. Rather than devaluing this approach by asserting that ``facts'' always trump experience, it might be more effective to help Gen Zers identify different epistemologies and understand how they produce different answers. 

In sum, Gen Zers' information sensibility practices do not make them universally more susceptible to misinformation. Generally, the effects of these practices appear determined by local social and family norms. The use of ``good enough'' reasoning, perhaps developed in response to information overload, may help Gen Zers resist the certainty promised by misinformation. Before overwriting these practices with new competencies, HCI or literacy-based interventions should first recognize the utility of information sensibility from the user perspective.

\subsection{Effects: period, age, cohort}\label{sec-effects}

Here, we hypothesize which findings have more broad implications, classifying them into period, age, and cohort effects (Table~\ref{tab-overview}). We make these suggestions based on previous qualitative studies (\eg \cite{wheeler2021belief,goldberg21}) and the literature reviewed in Section \ref{sec-rw}. These hypotheses provide avenues for future research based upon our findings. Period, age, and cohort effects are not always neatly categorizable, so we indicate in each section where gradients occur. 

\subsubsection{Period effect} 

Period effects are experienced by all age groups living at a particular moment in time. Some period effects are the result of a specific, discrete event; others may be caused by longer-term social, economic, or cultural conditions. We hypothesize that people of all ages seek social belonging. People across the population as a whole also currently experience information overload at unprecedented scale and speed (challenge) and use surrogate thinking to cope with it, relying on go-to sources (practice). 	

We propose encountering information in the wild rather than searching for it is a period effect (experienced by all ages) but is more pronounced among Gen Z. We theorize that this effect has a generational gradient rather than a sharp difference: younger generations are more likely to encounter most of their information rather than searching for it. Being pulled by a desire for social belonging more than a desire to ascertain accurate ``truths,'' we posit, is shared across the population but is more pronounced in adolescents (gradient age effect). Coupled with research into misinformation susceptibility in adult populations \cite{wheeler2021belief}, we also hypothesize that discomfort with uncertainty and desire for normativity predicts susceptibility to misinformation across the population, which we suggest testing in future research. 

\subsubsection{Age effect} 

Age effects are effects that people of a certain age experience, regardless of the time period they live in. Adolescence is a key example. We propose that a heightened fear of social error and dislike of misrecognition (challenges) are age effects more acutely experienced by adolescents. Adolescents also experience a stronger desire to orient themselves based upon their peers, and to develop a sense of self via information practices. 

More specifically, we noted an increased family influence on beliefs for 13--17 year olds, which we propose is an age effect, one that modifies the surrogates they select. As people age, they move away from family influence toward peer influence regardless of generation. The specific practices they use to do so, we propose, are cohort effects (to which we now turn). 

\subsubsection{Cohort effect} 

Cohort effects are effects experienced by specific generations (\eg Gen Z, Baby Boomers). Generational technology use has specific effects on information practices. For example, Gen Z practices are shaped by the media they use most as a cohort (\eg TikTok) at this particular, critical developmental stage. 

The specific information validation practices our participants used were therefore largely unique to Gen Z, due to the media ``waters'' they grew up swimming in. The affordances of the social media technologies they interacted with allowed for the crowdsourcing credibility practices they used. Generations coming of age before social media lacked these digital social cues altogether\footnote{We note this as both a cohort and period effect in Table~\ref{tab-overview} due to the familiarity that this generation has with social cues, though people of all ages \& cohorts may also use information from the comments as in~\cite{geeng2020fake}.}. 

Further, Gen Zers place higher trust in everyday experience (practice) rather than expertise, we posit, in part due to a widespread collapse of institutional trust as they were learning to evaluate information. This is a cohort effect, since it is dependent on the specific historical moment in which Gen Z came of age. Finally, Gen Zers are more risk-averse than other generations, in part due to stagnation of wages and a fear of getting publicly "canceled" \cite{katz2021gen}.

\section{Conclusion}
In this paper, we argued that Gen Zers practice \textit{information sensibility} when encountering and evaluating information online, interpreting it through the lens of their existing social influences. For Gen Zers---and, we hypothesize, for others---information processing is fundamentally a social practice. Our findings suggest that, like Gen Z's information sensibility practices, the solutions and strategies to address misinformation should be socially embedded. 

Information sensibility is complementary to information literacy, not a replacement for it. The descriptive reorientation provided by our work---and by the concept of information sensibility we introduce with it---opens up space for others to develop a new set of prescriptive competencies to foster with a goal of reducing vulnerability to misinformation. Such interventions should address how we can best support Gen Zers in meeting their own needs and building resilience to misinformation, given what we know about the social nature of information seeking, evaluating, and sharing. 

\begin{acks}
We are grateful to Vanessa Maturo for her early work on this project. We extend a very big thank you to Behzad Sarmadi, Rebekah Park, and Todd Carmody for their many efforts in support of this project. We would also like to thank Brad Chen, Cass Matthews, Crystal Lee, Francesca Tripodi, Hector Ouilhet Olmos, Ioana Literat, Jason Lipshin, Mariana Saavedra Espinosa, Meena Natarajan, Nidhi Hebbar, Paree Zarolia, Rachel A. Edson, Roberta R. Katz, Sam Wineburg, Serena Chao, Vishnupriya Das, Yasmin Green, Zach Cordier, and several other experts and colleagues for their support and ideas as we planned and conducted this research. Finally, we’d like to thank our participants for sharing their experiences with us, and our reviewers for helping to improve our paper.
\end{acks}

\bibliographystyle{ACM-Reference-Format}
\bibliography{genz}


\end{document}